\def\Bbb#1{{\bf #1}}

\def\fnote#1{\footnote}

\def\cwleftpar#1#2{\leftskip #1 \rightskip #2 plus 1fill}
\def\cwrightpar#1#2{\leftskip #1 plus 1fill \rightskip #2}
\def\cwcenterpar#1#2{\leftskip #1 plus 1fill \rightskip #2 plus 1fill}
\def\cwfullpar#1#2{\leftskip#1\rightskip#2}

\def\cwoutdent#1#2{\llap{\hbox to #1{#2 \hss}}\ignorespaces}
\def\cwparbegin#1#2#3#4#5{
	\ifcase #1 \cwleftpar{#2}{#3}
	\or \cwrightpar{#2}{#3}
	\or \cwcenterpar{#2}{#3}
	\else \cwfullpar{#2}{#3}\fi
	\ifcase #4 \baselineskip = 1.5\baselineskip
	\or \baselineskip = 2\baselineskip
	\or \baselineskip = 3\baselineskip
	\else \baselineskip = 1\baselineskip\fi
	\ifdim #5 > 0in \else \noindent \fi
	\noindent\ignorespaces}
\documentclass{article}
\begin{document}
\advance \vsize by -1\baselineskip
\def\makefootline{
\ifnum\pageno = 1{\vskip \baselineskip \vskip \baselineskip }\else{\vskip \baselineskip \noindent \folio                                  \par
}\fi}

 \vspace*{3ex}
\noindent {\Huge Generalized Doppler Effect in Spaces\\[1ex]
		 with a Transport along Paths}

 \vspace*{3ex}

\noindent Bozhidar Zakhariev Iliev
\fnote{0}{\noindent $^{\hbox{}}$ Permanent address:
Laboratory of Mathematical Modeling in Physics,
Institute for Nuclear Research and \mbox{Nuclear} Energy,
Bulgarian Academy of Sciences,
Boul.\ Tzarigradsko chauss\'ee~72, 1784 Sofia, Bulgaria\\
\indent E-mail address: bozho@inrne.bas.bg\\
\indent URL: http://theo.inrne.bas.bg/$\sim$bozho/}

 \vspace*{3ex}

{\bf \noindent Published: Communication JINR, E5-95-160, Dubna, 1995}\\[1ex]
\hphantom{\bf Published: }
http://www.arXiv.org e-Print archive No.~math-ph/0401033\\[2ex]

2003 PACS numbers: 04.90.+e, 45.90.+t\\[2ex]

\noindent
{\small
The \LaTeXe\ source file of this paper was produced by converting a
ChiWriter 3.16 source file into
ChiWriter 4.0 file and then converting the latter file into a
\LaTeX\ 2.09 source file, which was manually edited for correcting numerous
errors and for improving the appearance of the text.  As a result of this
procedure, some errors in the text may exist.
}\\[2ex]

	\begin{abstract}

An analog of the classical Doppler effect is investigated in spaces
(manifolds) whose tangent bundle is endowed with a transport along paths,
which, in particular, can be parallel one. The obtained results are valid
irrespectively to the particles mass, i.e.\ they hold for massless particles
(e.g. photons) as well as for massive ones.

	\end{abstract}\vspace{3ex}

 {\bf 1. INTRODUCTION}
\nopagebreak

\medskip
 The present paper continues the began in [1,2] applications of the theory of
transports along paths (in fibre bundles) [3] to the mechanics of material
point particles. Here is studied a phenomenon consisting in the comparison of
the (relative) energies of a material (massive or massless) point particle
with respect to two other arbitrary moving point particles (observers). An
evident special case of this problem is the well known Doppler effect [4,5].
When the mentioned transport along paths is linear and, in fact, only paths
without self-intersections are taken into account the above problem was
investigated in [6]. Here we closely follow [6] without supposing these
restrictions.

Sect. 2 contains the strict formulation of the problem of the present work and the derivation of the main results, which in Sect. 3 are applied to the general and special relativity. Sect. 4 closes the paper with some comments.

Below the needed for the following mathematical background is summarized.

All considerations in the present work are made in a (real) differentiable
manifold $M [7]$ whose tangent bundle $(T(M),\pi ,M)$ is endowed with a
transport along paths $I [3]$. Here $T(M):=\cup_{x\in M}T_{x}(M)$,
$T_{x}(M)$ being the tangent to $M$ space at $x\in M$ and $\pi :T(M)  \to M$
is such that $\pi (V):=x$ for $V\in T_{x}(M)$. Besides, the tangent bundle
$(T(M),\pi ,M)$ is supposed to be equipped also with a real bundle metric
$g$, i.e. $[8] a$ map $g:x\to g_{x}, x\in M$, where the maps
$g_{x}:T_{x}(M)\otimes T_{x}(M) \to {\Bbb R}$ are bilinear, nondegenerate and
symmetric. For brevity the defined by $g$ scalar products of $X,Y\in
T_{y}(M), y\in M$ will be denoted by a dot $(\cdot )$, i.e. $X\cdot
Y:=g_{y}(X,Y)$. The scalar square of $X$ will be written as $(X)^{2}$for it
has to be distinguished from the second component $X^{2}$of $X$ in some local
basis (in a case when $\dim(M)>1)$. As $g$ is not supposed to be positively
defined, $(X)^{2}$can take any real values.

By $J$ and $\gamma :J  \to M$ are denoted, respectively, an arbitrary real
interval and a path in M. If $\gamma $ is of class $C^{1}$, its tangent
vector is written as $\dot\gamma$.

 The transport along paths I and the bundle metric are supposed to be
consistent, i.e. I preserves the defined by $g$ scalar products of the
vectors (see $[3], eq. (2.9))$:
\[
  A\cdot B=(I^{\gamma }_{s\to t}A)\cdot (I^{\gamma }_{s\to t}B), A,B\in
T_{\gamma (s)}(M),\quad s,t\in J,
\]
where $I^{\gamma }_{s\to t}$ is the transport along $\gamma $ from $s$ to $t
[3]$.

For details concerning transports along paths the reader is referred to [3] and for the ones about relative mechanical quantities (such as velocity, momentum and energy) - to [1].

\medskip
\medskip
 {\bf 2. STATEMENT OF THE PROBLEM AND\\ GENERAL RESULTS}

\medskip
Let a material object (a point particle  which may be massless as well as
massive) be moving in $M$ along the path $\gamma :J\to M$ (its world line)
which is parameterized with $r\in $J. Let $\gamma $ intersects the paths
$x_{a}:J_{a}\to M, a=1,2$, representing the world lines of the particles 1
and 2, which we call observers; i.e. for some $r_{a}\in J$ and $s^{0}_{a}\in
J_{a}$, we have $\gamma (r_{a})=x_{a}(s^{0}_{a}), a=1,2$. If it is necessary,
the parameters $r\in J$ and $s_{a}\in J_{a}, a=1,2$ will be considered as
proper times of the corresponding particles. As a special case of this
construction we can point the case when the material object is emitted from
the first particle and/or is detected from the second one, or vice versa. In
particular, if the considered material object is a photon, then the last
situation realizes the classical Doppler effect [4,5].

We put the following problem. On the basis of the introduced in [1] concepts we wish to compare the relative energies of the material object with respect to the observers 1 and 2 at the points $\gamma (r_{1})=x_{1}(s^{0}_{1})$ and $\gamma (r_{2})=x_{2}(s^{0}_{2})$ respectively.

Let $p(r)$ be the momentum of the material object (the observed particle) at
$\gamma (r)$ and $V_{a}(s_{a})=$  $_{a}(s_{a}) a=1,2$ be the velocities of
the observers. For brevity we let
\[
 p_{a}:=p(r_{a}), V_{a}:=V_{a}(s^{0}_{a}),\quad a=1,2.\qquad (1)
\]

In accordance with the definition of a relative energy (see [1], sect. 4) we have to
compare the relative energies
\[
 E_{1}:=\epsilon ((V_{1})^{2})p_{1}\cdot V_{1}
\ \qquad
E_{2}:=\epsilon ((V_{2})^{2})p_{2}\cdot V_{2},\qquad (2)
\]
 of the observed particle with respect to the particles 1 and 2 at the points
$\gamma (r_{1})=x_{1}(s^{0}_{1})$ and $\gamma (r_{2})=x_{2}(s^{0}_{2})$
respectively. Here $\epsilon (\lambda ):=+1$ for $\lambda >0$ and $\epsilon
(\lambda ):=-1$ for $\lambda \le 0$.

  We introduce the quantities:
\[
 (V_{2})_{1}:=I^{\gamma }_{r_{2}\to r_1}V_{2},\qquad (3)
\]
\[
  \Delta p(r_{1},r_{2};\gamma )
:=p(r_{2}) - I^{\gamma }_{r_{1}\to r_2}p(r_{1})
=p_{2}- I^{\gamma }_{r_{1}\to r_2}p_{1}.\qquad (4)
\]

The quantity (4) is defined analogously to the relative momentum $(cf. [1]$, sect. 3), but it defines along $\gamma $ with the help of the transport along paths I the change of the momentum of the observed particle when it moves from $\gamma (r_{1})$ to $\gamma (r_{2})$.

  Now we shall prove that the quantity
 \[
  \Delta E_{21}:=\epsilon ((V_{2})^{2})\Delta p(r_{1},r_{2};\gamma )\cdot
V_{2}\qquad (5)
\]
  is closely connected with the change of the energy of the material object
along $\gamma $ with respect to the point $\gamma (r_{2})=x_{2}(s^{0}_{2})$,
at which its world line intersects the world line of the second particle,
when the object moves from $\gamma (r_{1})$ to $\gamma (r_{2})$ along $\gamma
$. In fact, using $(1), (4)$ and (5), we get
\[
 \Delta E_{21}=\epsilon ((V_{2})^{2})(p_{2}
- I^{\gamma }_{r_{1}\to r_2}p_{1})\cdot
V_{2}=E(r_{2},r_{2};\gamma ) - E(r_{1},r_{2};\gamma ).  \qquad (6)
\]
 Here
\[
 E(r,r_{2};\gamma )=\epsilon ((V_{2})^{2})(I^{\gamma }_{r\to r_{2}}p(r))\cdot
V_{2},\qquad (7)
\]
 as a consequence of the considerations of [1], is the
defined along $\gamma $ by means of the transport along paths relative energy
of the observed particle when it is situated at $\gamma (r), r\in [r^\prime
,r^{\prime\prime}]$ with respect to the second particle when it is situated
at $\gamma (r_{2})=x_{2}(s^{0}_{2})$.

 From one hand, putting $r=r_{1}$in (7) and (due to the consistency between
the metric and the transport along paths) applying to the both multiplies
$I^{\gamma }_{r_{2}\to r_1}$ and, from the other hand, letting $r=r_{2}$in
(7) and using (2), we, respectively, get

 \[
  E(r_{1},r_{2};\gamma )=\epsilon ((V_{2})^{2})p_{1}\cdot (V_{2})_{1},
\quad E(r_{2},r_{2};\gamma )=E_{2}.\qquad (8)
\]

 So, from (6), we find
\[
  E_{2}=\Delta E_{21}+ \epsilon ((V_{2})^{2})p_{1}\cdot (V_{2})_{1}.\qquad
(9)
\]

 Further, supposing $(V_{1})^{2}\neq 0$, which is interpreted as a movement
of the first observer with a velocity different from (less than) the one of
light in vacuum, we shall express $p_{1}\cdot (V_{2})_{1}$through $E_{1}$.
Representing $(V_{2})_{1}$in the form $(V_{2})_{1}=(V_{2})^{\parallel }_{1}+
(V_{2})^{\bot}_{1}$, where the longitudinal $(V_{2})^{\parallel }_{1}$ and the
transversal $(V_{2})^{\bot}_{1}$components with respect to $V_{1}$ are given
by
\[
 (V_{2})^{\parallel }_{1}:=V_{1}(V_{1}\cdot (V_{2})_{1})/(V_{1})^{2}
 \qquad
((V_{2})^{\parallel }_{1}\cdot V_{1}=(V_{2})_{1}\cdot V_{1}),\qquad (10a)
\]
\[
(V_{2})^{\bot}_{1}:=(V_{2})_{1}- (V_{2})^{\parallel }_{1}
 \qquad
((V_{2})^{\bot}_{1}\cdot V_{1}=(V_{2})^{\bot}_{1}\cdot (V_{2})^{\parallel
}_{1}=0),\qquad (10b)
\]
 we obtain:
\[
 p_{1}\cdot (V_{2})_{1}= p_{1}\cdot (V_{2})^{\bot}_{1}+ p_{1}\cdot
(V_{2})^{\parallel }_{1}= p_{1}\cdot (V_{2})^{\bot}_{1}
\]
\[
 + (p_{1}\cdot V_{1})(V_{1}\cdot (V_{2})_{1})/(V_{1})^{2}= p_{1}\cdot
(V_{2})^{\bot}_{1}
\]
\[
 + \epsilon ((V_{1})^{2})E_{1}[(((V_{2})_{1})^{2}-
((V_{2})^{\bot}_{1})^{2})/(V_{1})^{2}]^{1/2},\qquad (11)
\]
 where we have used (2) and the equality $(V_{1}\cdot
(V_{2})_{1})/(V_{1})^{2}
= [((V_{2})^{\parallel }_{1})^{2}/(V_{1})^{2}]^{1/2}
= [(((V_{2})_{1})^{2}- ((V_{2})^{\bot}_{1})^{2})/(V_{1})^{2}]^{1/2}$
(the square root sign is uniquely defined by
$(V_{1}\cdot (V_{2})_{1})/(V_{1})^{2}|_{(V_{2})_1=V_1}=+1)$,
which follows from $((V_{2})^{\parallel }_{1})^{2}=(V_{1}\cdot
(V_{2})_{1})^{2}/(V_{1})^{2}$ and $((V_{2})^{\bot}_{1})^{2}=
((V_{2})_{1})^{2}-((V_{2})^{\parallel }_{1})^{2})$.

 Substituting (11) into (9), we get:
\[
  E_{2}=\Delta E_{21}+\epsilon ((V_{1})^{2})\epsilon
((V_{2})^{2})E_{1}
[(((V_{2})_{1})^{2}-((V_{2})^{\bot}_{1})^{2})/(V_{1})^{2}]^{1/2}
\]
\[
+ \epsilon ((V_{2})^{2})p_{1}\cdot (V_{2})^{\bot}_{1}.\qquad (12)
\]

This formula is the answer of the stated above problem and it expresses the "generalized" Doppler's effect for the considered process.

 Now we will put the last term of (12) into a slightly different form $(cf.
[4])$.

 Let the vector $N_{1}\in T_{\gamma (r_{1}}(M)$ be defined in the following
way. If $p_{1}$and $V_{1}$are not collinear, then $N_{1}$is coplanar with
them, i.e. $N_{1}=aV_{1}+bp_{1}$, for some $a,b\in {\Bbb R}$, and satisfies
the conditions
 \[
N_{1}\cdot V_{1}=0,\qquad (13a)
\]
\[
 (N_{1})^{2}=N_{1}\cdot N_{1}=\epsilon
((p_{1})^{2}-(E_{1})^{2}/(V_{1})^{2}),\qquad (13b)
\]
\[
 N_{1}\cdot p_{1}<0.\qquad (13c)
\]
In this case $N_{1}$is uniquely defined and its connection with $p_{1}$and
$V_{1}$may be written, for example, as
\[
 p_{1}=V_{1}(V_{1}\cdot p_{1})/(V_{1})^{2}
\]
\[
 - N_{1}\epsilon ((p_{1})^{2}-(E_{1})^{2}/(V_{1})^{2})\mid
(p_{1})^{2}-(E_{1})^{2}/(V_{1})^{2}\mid ^{1/2},\qquad (14a)
\]
 where $\mid \lambda \mid :=\lambda \epsilon (\lambda )$ is the absolute
value of $\lambda \in {\Bbb R}. ($Note that
$(p_{1})^{2}-(E_{1})^{2}/(V_{1})^{2}=0$ if and only if $p_{1}$ and $V_{1}$ are
collinear; see (1).) If $p_{1}$ and $V_{1}$ are collinear, i.e. $p_{1}=\lambda
V_{1}$for some $\lambda \in {\Bbb R} ($which, because of $(V_{1})^{2}\neq 0$,
is given by $\lambda =p_{1}\cdot V_{1}/(V_{1})^{2})$, then we put $N_{1}=0$:
\[
 N_{1}=0\quad  for\ p_{1}=\lambda V_{1}.\qquad (14b)
\]
 (Note that due to (14a) from $N_{1}=0$ follows $p_{1}=\lambda V_{1}$, which
is equivalent to $(p_{1})^{2}-(E_{1})^{2}/(V_{1})^{2}=0.)$

Hence the defined by (14) vector $N_{1}\in T_{\gamma (r_{1}}(M)$ is orthogonal to a path $x_{1}$at the point $\gamma (r_{1})=x_{1}(s^{0}_{1}) (N_{1}\cdot V_{1}=0)$ and if $p_{1}$and $V_{1}$are not collinear, then it is a unit vector "pointing from $x_{1}$to $x_{2}" ((N_{1})^{2}=\epsilon ((p_{1})^{2}-(E_{1})^{2}/(V_{1})^{2}))$.

By definition the recession speed $\omega_{21}$ of the particle 2 from the
particle 1 is the projection of $(V_{2})_{1}$on $N_{1}($see [4]), i.e.
\[
\omega_{21}=(V_{2})_{1}\cdot N_{1}=(V_{2})^{\bot}_{1}\cdot N_{1}.\qquad (15)
\]

 If $p_{1}$and $V_{1}$are collinear, then $ \omega _{21}=0$.

 From $(14), (15)$ and (10b), we obtain
\[
  p_{1}\cdot (V_{2})^{\bot}_{1}=-  _{21}\epsilon
((p_{1})^{2}-(E_{1})^{2}/(V_{1})^{2})\mid
(p_{1})^{2}-(E_{1})^{2}/(V_{1})^{2}\mid ^{1/2}.  \qquad (16)
\]

Substituting this expression into (12), we find the looked for connection
between the relative energies $E_{1}$and $E_{2}$in the form
\[
 E_{2}=\Delta E_{21}+ \epsilon ((V_{1})^{2})\epsilon
((V_{2})^{2})E_{1}
[(((V_{2})_{1})^{2}-((V_{2})^{\bot}_{1})^{2})/(V_{1})^{2}]^{1/2}
\]
\[
 - \epsilon ((V_{2})^{2})\epsilon ((p_{1})^{2}-(E_{1})^{2}/(V_{1})^{2})
_{21}\mid (p_{1})^{2}-(E_{1})^{2}/(V_{1})^{2}\mid ^{1/2}.  \qquad (17)
\]

Namely this formula gives the Doppler's effect in terms of particles and
their energies in the considered process, which can be called a "generalized
Doppler effect". The corresponding to it "red shift" [4] is given by the
equality
\[
 (E_{2}-E_{1})/E_{2}=1 - \bigl\{ \Delta E_{21}/E_{1}+ \epsilon
((V_{1})^{2})\epsilon ((V_{2})^{2})[(((V_{2})_{1})^{2}
\]
\[
  -((V_{2})^{\bot}_{1})^{2})/(V_{1})^{2}]^{1/2}- \epsilon
((V_{2})^{2})\epsilon ((p_{1})^{2}-(E_{1})^{2}/(V_{1})^{2})
\]
\[
\times
 \epsilon (E_{1})  _{21}\mid (p_{1})^{2}/(E_{1})^{2}-1/(V_{1})^{2}\mid
^{1/2} \bigr\} ^{-1}.\qquad (18)
\]

\medskip
\medskip
 {\bf 3. EXAMPLES: GENERAL AND SPECIAL RELATIVITY}

\medskip
In this section are consider the applications of the obtained in Sect. 2 general results to the cases of the Doppler's effect in general relativity and the "generalized Doppler effect" (for constant velocities) in special relativity.

The Doppler effect in general relativity is investigated, e.g. in $[4], ch$. III, section 7, where the condition (13c) is not stated explicitly but it is used in the calculation, consist in the following. At the point $x_{2}(s^{0}_{2})$ is emitted a photon which moves along the isotropic geodesic path $\gamma $ to the point $x_{1}(s^{0}_{1})$, where it is detected. The problem is to be compared the detection and emission energies $E_{1}$and $E_{2}$respectively.

Choosing the manifold $M$ to be the space-time of general relativity with
signature $(-+++)$ and $I^{\gamma }$to be a parallel transport along $\gamma
$, we find (see section 2 and $[1]): (V_{1})^{2}=(V_{2})^{2}=-c^{2}(c$ is the
velocity of light in vacuum$), (p_{1})^{2}=(p_{2})^{2}=0, E_{1}>0, \Delta
p(r_{1},r_{2};\gamma )\equiv 0, \Delta E_{21}=0$ and
$N_{1}=-[p_{1}+V_{1}(V_{1}\cdot p_{1})]/E_{1}$. So, (17) and (18) take
respectively the form:
\[
 E_{2}=E_{1}\omega_{21}+(1+((V_{2})^{\bot}_{1})^{2}/c^{2})^{1/2}],\qquad (19)
\]
\[
(E_{2}-E_{1})/E_{2}=1-[\omega
_{21}+(1+((V_{2})^{\bot}_{1})^{2}/c^{2})^{1/2}]^{-1}.\qquad (20)
\]

 By using a local orthonormal basis this result is derived in $[4], ch$. III,
section 7, where $E_{1}, E_{2}, \omega_{21}$ and $(V_{2})^{\bot}_{1}$ are
denoted as $E, E^\prime ,  \omega_{R}$ and $\omega^{2}$ respectively and,
besides, there are used units in which $c=1$.

In  the case of special relativity $M$ is the Minkowski's space-time and
$I^{\gamma }$is also a parallel transport along $\gamma $. Let the considered
particles be moving with constant 3-velocities ${\bf v}, {\bf v}_{1}$and
${\bf v}_{2}$with respect to a given frame of reference, i.e. we have $\gamma
(r)=y+(ct,t{\bf v})$ and $x_{a}(s_{a})=y_{a}+(ct,t{\bf v}_{a}), a=1,2$, where
$y,y_{1},y_{2}\in M$ are fixed, $t$ is the time in that frame, $r=t$ for
$\mid {\bf v}\mid =c$ and $r=t(1-{\bf v}^{2}/c^{2})^{1/2}$for $\mid {\bf
v}\mid <c$, and $s_{a}=t(1-{\bf v}^{2}_{a}/c^{2})^{1/2}, a=1,2$ are the
corresponding proper times (cf.\ [2,4,5]). (Of course, we suppose that
$\gamma $ intersects $x_{1}$and $x_{2}.)$

From here we find the momenta of the observed particle as $p(r)=
=p_{1}=p_{2}=\mu c(1,{\bf v}/c)$, where $\mu :=E/c$ for $\mid {\bf v}\mid =c
(E$ is the particle's energy in the given frame) and $\mu :=m(1-{\bf
v}^{2}/c^{2})^{-1/2}$for $\mid {\bf v}\mid <c (m$ is the particle's rest
mass$), {\bf v}_{a}=c(1,{\bf v}/c)(1-{\bf v}^{2}_{a}/c^{2})^{-1/2}, a=1,2,
(V_{2})_{1}=V_{2}, \Delta p(r_{1},r_{2};\gamma )\equiv 0$ and $\Delta
E_{21}=0$. The computation of the remaining quantities concerning the
phenomenon under consideration is simple but long; that is why we present
here only the final result:
\[
E_{2} = E_{1}
\Bigl(\frac{1-{\bf v}_{1}^{2}/c^2}{1-{\bf v}_{2}^{2}/c^2}
\Bigr)^{1/2}
\cdot
\frac{1-{\bf v}_{2}\cdot{\bf v}/c^2}
     {1-{\bf v}_{1}\cdot{\bf v}/c^2} ,
 \qquad (21)
\]
which also may easily be obtain from the expressions $E_{a}= =\epsilon
((V_{a})^{2})p_{a}\cdot V_{a}=\mu c^{2}(1-{\bf v}_{a}\cdot {\bf
v}/c^{2})(1-{\bf v}^{2}_{a}/c^{2})^{-1/2}, a=1,2$.

In the case of the "usual" Doppler effect [5] we have a photon $({\bf v}=c{\bf n}, {\bf n}^{2}=1)$ emitted with energy $E_{2}=E_{0}$and detected with energy
$E_{1}=E$, which according to (21) is

\[
E = E_{0}
\Bigl(\frac{1-{\bf v}_{2}^{2}/c^2}{1-{\bf v}_{1}^{2}/c^2}
\Bigr)^{1/2}
\cdot
\frac{1-{\bf v}_{2}\cdot{\bf n}/c^2}
     {1-{\bf v}_{1}\cdot{\bf n}/c^2} ,
 \qquad (22)
\]

 Let us note that when ${\bf v}_{1}={\bf 0}$, the formulae (21) and (22) may
be derived also as a corollary from the definition of relative energy (see
[1], sect. 4). This result is in agreement with the considered in [5] Doppler
effect in special relativity (in terms of frequencies; cf. the quantum
relation $E=h\nu , h$ being the Planck's constant).

\medskip
\medskip
 {\bf 4. CONCLUDING REMARKS}

\medskip
We want to emphasize that the basic result of this work is given by the equality (17) the main difference of which from the usual Doppler effect (e.g. in general relativity; see (19) or [3]) is the existence in it of, generally said, nonvanishing term $\Delta E_{21}$, defined by $(5). A$ feature of the equation (17) is its validity for particles with arbitrary, zero or nonzero, masses, i.e. our results do not depend on the mass of the investigated particle.

If $\Delta E_{21}=0$, then $eq. (17)$ takes  a form similar to the classical
one (in the case of arbitrary mass $(cf. [4]))$. Due to $(5) a$ sufficient
condition for this is
\[
 \Delta p(r_{1},r_{2};\gamma )=0,\qquad (23)
\]
i.e. (see (4))
\[
 p(r_{2})=I^{\gamma }_{r_{1}\to r_2}p(r_{1})\qquad (23^\prime )
\]
 which means that the momentum $p(r_{1})$ is (I-)transported by means of the
transport along paths I from the point $\gamma (r_{1})$ to the point $\gamma
(r_{2}) (cf. [3], eq. (2.4))$.

If a point particle is moving along the path $\gamma :J   \to M$, then it is
natural to call it a {\it free particle} (with respect to the transport along
paths I) if $(23^\prime )$ holds for every $r_{1},r_{2}\in J$, i.e. if its
momentum is I-transported along its world line $\gamma  (cf. [3]$, definition
2.2). In particular, if $I^{\gamma }$is a parallel transport along $\gamma
($generated by a linear connection), then this definition of a free particle
coincides with the one in [4], p. 110, given therein as a special case of the
geodesic hypothesis. In our case, the corresponding generalization of the
geodesic hypothesis states that the world line $\gamma :J  \to M$ of a tree
(with respect to I) particle is an I-path (see [2], definition 2.2), i.e.
\[
\dot\gamma(t)=I^{\gamma }_{s\to t}\dot\gamma(s),
\quad s,t\in J,\qquad (24)
\]
and besides
\[
 p(s)=\mu (s;\gamma )\dot(s)\qquad (25)
\]
 for some scalar function $\mu  ($identified with the particles rest mass if
it is not zero).

 If the transport along paths is linear (see e.g. $[3], eq. (2.8))$, then
substituting (25) into $(23^\prime )$ and comparing the result with (24), we
get $(cf. [4]$, p. $110, eq. (9))$
\[
 \mu (s;\gamma )=const.  \qquad (26)
\]

So, the mass parameter $\mu  ($the rest mass if it is not zero) of a free particle is constant if the used transport along paths is linear.

It should also be noted, that as for free massive particles with a rest mass $m eq. (25)$ holds (by definition) for $\mu (s;\gamma )=m$, then for linear transports along paths the ("generalized geodesic") hypothesis (24) is a consequence of the condition $(23^\prime )$.

  At the end we want to mention the equality $(cf. (4))$
\[
  \Delta p(r_{2},r_{1};\gamma )
=I^{\gamma }_{r_{2}\to r_1}\Delta p(r_{1},r_{2};\gamma )\qquad (27)
\]
 for a linear transport along paths I. Hence, in this case $eq. (5)$ can
equivalently (due to the consistency of I and the metric) be written as
\[
 \Delta E_{21}:=\epsilon ((V_{2})^{2})\Delta p(r_{2},r_{1};\gamma )\cdot
(V_{2})_{1}.\qquad (28)
\]

\medskip
\medskip
 {\bf ACKNOWLEDGEMENT}

\medskip
This research was partially supported by the Fund for Scientific Research of Bulgaria under contract Grant No. $F 103$.

\medskip
\medskip
 {\bf REFERENCES}

\medskip
1.  Iliev B.Z., Relative mechanical quantities in spaces with a transport along paths, JINR Communication $E2-94-188$, Dubna, 1994.\par
2.  Iliev B.Z., Deviation equations in spaces with a transport along paths, JINR Communication $E2-94-40$, Dubna, 1994.\par
3.  Iliev B.Z., Transports along paths in fibre bundles General theory, Communication JINR, $E5-93-299$, Dubna, 1993.\par
4.  Synge J.L., Relativity: The general theory, North-Holland Publ. Co., Amsterdam, 1960.\par
5.  Moller C., The Theory of Relativity, $2^d$ ed., Clarendon Press, Oxford, 1972\par
6.  Iliev B.Z., On one generalization of the Doppler effect in spaces with general linear transport, Proceedings of the $3^d$ international seminar "Gravitational energy and gravitational waves", JINR, Dubna, $19-21$ may 1990, Dubna, 1991.
\par
7.  Hawking S.W., G.F.R. Ellis, The large scale structure of space-time, Cambridge Univ. Press, Cambridge, 1973.\par
8.  Dubrovin B.A., S.P. Novikov, A.T. Fomenko, Modern geometry, Nauka, Moscow, 1979 (In Russian).

\newpage

\medskip
\medskip
\medskip
\noindent Iliev B. Z.\\[5ex]

 \noindent Generalized Doppler Effect\\
 in Spaces with a Transport along Paths

\medskip
\medskip
\medskip
An analog of the classical Doppler effect is investigated in spaces
(manifolds) whose tangent bundle is endowed with a transport along paths,
which, in particular, can be parallel one. The obtained results are valid
irrespectively to the particles mass, i.e.\ they hold for massless particles
(e.g. photons) as well as for massive ones.
\\[5ex]

\medskip
\medskip
The investigation has been performed at the Laboratory of Computing
Techniques and Automation, JINR.

\end{document}